\documentclass[aps,reprint,amsmath, amssymb,groupedaddress,floatfix]{revtex4-1}
\usepackage{natbib}
\usepackage{graphicx}
\usepackage{bm}
\usepackage{color,soul}
\usepackage{hyperref}
\usepackage{float}
\usepackage{csquotes}
\usepackage{color,soul}
\usepackage{hyperref}
\hypersetup{colorlinks=true, urlcolor=blue, citecolor=blue}

\begin{document}

\preprint{APS}
\title{Giant thermopower and power factor in magic angle twisted bilayer graphene at low temperature}

\author{S. S. Kubakaddi}
\email{sskubakaddi@gmail.com}
\affiliation{
 Department of Physics, K. L. E. Technological University, Hubballi-580031, Karnataka, India
}

\date{\today}
\begin{abstract}

The in-plane phonon-drag thermopower $S^g$, diffusion thermopower $S^d$ and the  power factor $PF$ are theoretically investigated in twisted bilayer graphene (tBLG) as a function of twist angle $\theta$, temperature $T$ and electron density $n_s$ in the region of low $T$ (1-20 K). As $\theta$ approaches magic angle $\theta_m$, the $S^g$ and $S^d$ are found to be strongly enhanced, which is manifestation of great suppression of Fermi velocity ${\nu_F}^*$ of electrons in moire flat band near $\theta_m$. This enhancement decreases with increasing $\theta$ and $T$. In the Bloch- Gruneisen (BG) regime, it is found that $S^g \sim  {\nu_F}^{* -2}$, $T^3$ and ${n_s}^{-1/2}$. As $T$ increases, the exponent $\delta$ in $S^g \sim T^\delta$, changes from 3 to nearly zero and a maximum $S^g$ value of $\sim$ 10 mV/K  at $\sim$ 20 K is estimated. $S^g$ is larger (smaller) for smaller $n_s$ in low (high) temperature region. On the other hand, $S^d$, taken to be governed by Mott formula, $\sim {\nu_F}^{* -1}$, $T$ and ${n_s}^{-1/2}$ and $S^d<<S^g$  for $T > \sim$ 2 K. The power factor $PF$ is also found to be strongly $\theta$ dependent and very much enhanced. Consequently, possibility of a giant figure of merit is discussed.In tBLG, $\theta$ acts as a strong tuning
parameter of both $S^g$ and $S^d$ and $PF$ in addition to $T$ and $n_s$. Our results are qualitatively compared with the measured out-of-plane thermopower in tBLG.
\end{abstract}
\maketitle

\section{ INTRODUCTION}

The recent remarkable and exciting experimental discoveries in twisted bilayer graphene (tBLG), particularly the existence of correlated insulating phases and superconductivity at low temperatures, and a highly resistive linear-in-temperature resistivity $\rho$ at high temperature, have created tremendous interest in the study of electronic properties of this material \cite{1,2,3,4,5,6,7,8,9,10,11,12,13,14,15}. A small twist angle $\theta$ between the two layers modifies the electronic structure in to moire flat band with $\theta$ dependent suppressed  Fermi velocity ${\nu_F}^* \equiv {\nu_F}^*$($\theta$), which is very much smaller than the bare Fermi velocity ${\nu_F}$ in monolayer graphene, and becomes vanishingly small for $\theta$ in the neighborhood of magic angle $\theta_m$ leading to large density of states \cite{10, 13, 14, 16}. The $\theta$ near $\theta_m$, acts as one of the tunable parameters in limiting the electronic properties of tBLG, apart from the carrier density $n_s$ and temperature $T$ \cite{1, 2, 4, 9, 10, 11,13}. 

The measurements of in-plane electrical resistivity $\rho$ in tBLG, for $\theta$ closer to $\theta_m$, show that $\rho$ scales linear-in-$T$, strongly increasing with decreasing twist angle and enhanced by about more than three orders of magnitude over that observed in monolayer graphene (MLG) \cite{5,9,11}. This linearity is observed, in some samples, unbelievably down to 0.5 K \cite{5,11}, just before the samples become superconducting, and in some other samples it persists up to $\sim$ 4 - 6 K \cite{5,9,11}. The linear-in-$T$ behavior of $\rho$ is generic of electron scattering by acoustic phonons at high T. The acoustic phonon limited electrical resistivity $\rho_p$ in tBLG is theoretically investigated assuming the linear dispersion for the Dirac fermions \cite{9,10,13}. It is found to be strongly $\theta$ dependent, largely enhanced in magnitude and linear-in-$T$ for the temperatures above $\sim$ 5 K and explaining the experimental data very well. The strong $\theta$ dependence and enhancement of $\rho_p$ is attributed to the increase in the electron-acoustic phonon (el-ap) scattering due to the suppression of ${\nu_F}^*$ induced by the moiré flat band.

Very recently the hot electron power loss $P$, another interesting property, which involves only el-ap interaction, has been theoretically investigated in tBLG by the present author \cite{15}. The profound effect of $\theta$, near $\theta_m$, on $P$ is found to enhance it by $\sim$ 2-3 orders of magnitude, attributing to the reduced ${\nu_F}^*$, complimenting the measurements of $\rho$.

The thermoelectric (TE) power, of any realized new material, is an important and interesting transport property for the study because of its sensitivity to electronic structure and scattering mechanisms, besides applications in TE devices (power generators and refrigerators). It is also known to provide information complementary to that of electrical resistivity $\rho$ (or conductivity $\sigma$) and has been a useful tool for probing carrier transport. The potential of a material for TE applications is determined by its figure of merit, $Z = S^2\sigma/\kappa$, where $S$ is the thermopower i.e. Seebeck coefficient, $\sigma$ is the electrical conductivity, and $\kappa$ is the thermal conductivity of the material. Materials, therefore, with an enhanced power factor $PF =  S^2\sigma$ and reduced $\kappa$, are suitable candidates for efficient TE devices \cite{17,18}. Thermopower $S$ defined by $E= S\nabla T$, where $E$ is  an electric field set up due to the temperature gradient $\nabla T$, comprises of  two components $S = S^d + S^g$: the diffusion thermopower $S^d$ arising due to the diffusion of electrons and the phonon-drag thermopower $S^g$ due to the momentum transfer from the  phonon wind to electrons because of the electron-phonon coupling. Since both $S^d$ and $S^g$ depend upon Fermi energy and hence the Fermi velocity and electron density, apart from temperature, we expect them to be strongly influenced by the  twist angle closer to $\theta_m$ in tBLG. More importantly, as we have seen the strong enhancement in the el-ap coupling dependent $\rho$ \cite{9,10,13} and $P$ \cite{15}, we expect the strong influence of $\theta$ on $S^g$, which depends only upon el-ap interaction. Moreover, the power factor $PF$ is also expected to be largely governed by $\theta$ in tBLG as both $\rho$ and $S$ are influenced. Consequently, the twist angle may act as an additional important tunable parameter for $Z$ in tBLG.

Thermopower has been studied in detail experimentally and theoretically in conventional two-dimensional electron gas (2DEG) (see Reviews \cite{19,20,21}) and in monolayer and conventional bilayer graphene (BLG) (see Review \cite{22}). In MLG, theories of phonon-drag thermopower \cite{23} and diffusion thermopower \cite {24} have been developed and experimentally investigated by different groups \cite{25,26,27,28}. In conventional BLG, phonon-drag thermopower and diffusion thermopower are studied theoretically \cite{29, 30} and compared with the experimental results \cite{31}.

Recently, thermoelectric transport has also been investigated, for specific values of $\theta$, in tBLG (across the plane) \cite{32_1, 33_1}, twisted armchair graphene nanoribbons (tAGNR) \cite{34_1} and twisted bilayer graphene nanoribbons
junction (tBGNRJ) \cite{35_1}. In their very recent work in tBLG, Mahapatra et al \cite{33_1} have reported out-of-plane thermoelectric measurements across the van der Waals gap
and shown that at large $\theta$ ($\sim$ 12.5$^\circ$) the thermopower, for $T=$20-300 K, is entirely driven by a novel phonon-drag effect at subnanometer scale, while at low twist angles ($\theta < 6^\circ$) the diffusion component of the thermopower is recovered that can be described by the Mott relation. These authors have used the phonon-drag theory of semiconductor quantum wells \cite{36_1}, with the layer breathing mode (LBM) of phonons of quadratic dispersion, to interpret their observations. In tBGNRJ from the first principle calculations, the enhanced thermoelectric performance with $ZT=$ (2-6.1) at 300 K, for optimized large $\theta$ ($\sim$ 21.8 $^\circ$ ), is interpreted as the combination of the reduced $\kappa$,
enhanced $S$ and $\sigma$ \cite{35_1}. We emphasize that the in-plane thermoelectric transport in tBLG is yet to be explored.

In the present work, we theoretically investigate the in-plane thermoelectric transport properties namely thermopower and power factor of electrons in moiré flat band
in tBLG at low temperature for small twist angles closer to magic angle. The effect of small twist angle on the in-plane phonon-drag and diffusion thermopower and the power factor
in tBLG is explored. In particular, study of $S^g$ is expected to compliment the enhancement of el-ap interaction found in the in-plane resistivity measurements. $S^g$, $S^d$ and $PF$ are  studied as a function of twist angle, temperature and electron density. We show that the twist angle $\theta$ acts as one of the strong tunable parameters of thermopower and power factor.\\

\section {Theoretical model}

As described in Ref \cite{15}, we assume the electron energy spectrum in moiré flat band to be given by the Dirac linear dispersion with the density of states $D(E_k) = (gE_k)/
[2\pi(\hbar {\nu_F}^*)^2]$, where $E_k = \hbar {\nu_F}^*|k|$ and the degeneracy $g = g_s g_v g_l$, with $g_s$, $g_v$ and $g_l$ being, respectively, spin, valley and
layer degeneracies each with the value of 2. The assumption of linear energy dispersion, in this early stage of the development, makes our theory limited to the carrier density $n_s \leq 10^{12} cm^{-2}$ with Fermi energy little away from the Dirac point \cite{10,13}. Das Sarma et al \cite{13} have shown that the twist angle dependence of ${\nu_F}^* (\theta$) can  be very well approximated by  ${\nu_F}^* (\theta)  \approx 0.5 \mid \theta - \theta_m\mid {\nu_F}$, which clearly indicates that effect of  $\theta$ is very large near   $\theta_m$. 

\subsection{Phonon-drag thermopower}	
We calculate the phonon-drag thermopower in tBLG following the well established Q-approach theory of Cantrell-Butcher (CB) given for conventional 2DEG \cite{32}, which is also applied in MLG by the present author \cite{23}. In this approach, for an in-plane electric field $E$ and temperature gradient $\nabla T$, the coupled Boltzmann equations for electrons and phonons are solved for electron distribution $f_\textbf{k}$ to obtain phonon-drag current density $J^g$ set by transfer of phonon momentum to electrons due to the electron-phonon coupling. In an open circuit condition, an electric field $E = S^g \nabla T$  is developed to stop this current density. 

Considering the interaction of acoustic phonons of energy $\hbar w_q$ and wave vector $q$ with the Dirac electrons in the moiré miniband tBLG \cite{10,15}, and following Refs. \cite{23,32}, we obtain

\begin{widetext}
\begin{align}
    S^g = - \frac{geA}{4\pi^2\sigma k_B T^2\hbar^5 \nu_s {\nu}_F^*} \int_0^\infty d(\hbar\omega_q) \int_\gamma^\infty dE_k (\hbar\omega_q)^2 \frac{{\mid M(q) \mid}^2 \tau_p \tau(E_k)} {{[1-(\gamma/E_k)^2]}^{1/2}}   \times \\ \nonumber
    [\{1+(\hbar \omega_q/E_k)\}\{1+(\hbar \omega_q/2E_k)\}]f(E_k)[1-f(E_k+\hbar \omega_q)]N_q ,
\end{align}
\end{widetext}
where $e$ is the electron charge, $A$ is the area of the tBLG, $\gamma =(E_q/2), E_q=\hbar {\nu_F}^*q$, ${\mid M(q) \mid}^2$ is the el-ap matrix element, $f(E_{\textbf{k}})= {[exp{(E_{\textbf{k}}-\mu)/k_BT} +1]}^{-1}$ is the Fermi-Dirac distribution with chemical potential $\mu$, $\tau(E_{\textbf{k}})$ and $\tau_p$ are electron and phonon momentum relaxation times, respectively, $N_{\textbf{q}}= {[exp (\hbar \omega_{\textbf{q}}/k_BT)- 1]}^{-1}$ is the phonon distribution function, $\omega_{\textbf{q}} = \nu_sq$ and $\nu_s$ is the acoustic phonon velocity.

The experimental observations of electrical
conductivity \cite{33} and power loss \cite{34,35} in MLG and linear-in-$T$ resistivity in tBLG \cite{10,13} are very well explained by the Dirac electron interaction with only LA phonons, without
screening. Hence, we use the corresponding el-ap interaction matrix element, given in Refs. \cite{10,13,15} for tBLG, in Eq.(1). Taking $\sigma = e^2 {\nu_F}^{*2}D({E_F}^*)\tau({E_F}^*)/2$ $[= 2e^2{E_F}^* \tau({E_F}^*)/\pi \hbar^2]$ \cite{10}, at Fermi energy, as approximated in Refs. \cite{23,32}, and the phonon relaxation time in  the boundary scattering regime $\tau_p = \Lambda / \nu_s$ \cite{23,32,36,37} with the  phonon mean free path  $\Lambda$ being smallest dimension of the sample, an expression for $S^g$ is found to be
\begin{widetext}
\begin{align}
    S^g & = - \frac{gD^2F(\theta)\Lambda}{16\pi e \rho_m k_BT^2\hbar^3{\nu_s}^4 {\nu_F}^*E_F^*} \int_0^\infty d(\hbar\omega_q) \int_\gamma^\infty dE_k (\hbar\omega_q)^3 \times \nonumber\\
    &{[1-(\gamma/E_k)^2]}^{1/2}  \times [\{1+(\hbar \omega_q/E_k)\}\{1+(\hbar \omega_q/2E_k)\}]f(E_k)[1-f(E_k+\hbar \omega_q)]N_q ,
\end{align}
\end{widetext}
where $D$ is the first-order acoustic deformation potential coupling constant and
$\rho_m$ is the areal mass density,$E_F^* = \hbar{\nu_F}^*k_F^*$, $k_F^* = (\pi n_s/2)^{1/2}$, and $F(\theta)$ is the form factor due to moiré wave function, modifying the el-ap interaction matrix element in tBLG \cite{10} as compared with the MLG.

At very low temperature, $q << 2k_F$ and $\hbar \omega_{\textbf{q}} \approx k_BT$ and the Bloch-Gruneisen (BG) regime is characterized by the temperature $T_{BG} = 2\hbar \nu_sk_F^*/k_B = 38 \sqrt{N}  K$, with $\nu_s$ = 2.0×10$^6$ cm/s and  $n_s$ = $N \times$ 10$^{12}$ cm$^{- 2}$. Additionally, we believe that, besides $q << 2k_F^*$, the condition $\hbar \omega_q << E_F^*$ needs to satisfied because of the very much suppressed ${\nu_F}^*$ which is $\theta$ dependent. Consequently, we anticipate the influence of $\theta$ on the temperature range of validity of BG regime power law obtained below. In tBLG, the $T_{BG}$ is smaller by a factor of $\sqrt{2}$ compared to MLG. For $T << T_{BG}$, $q\rightarrow 0$, the $S^g$  is given by the power law

\begin{equation}
    S^g = -S^{g0}T^{\delta}, \hspace{1cm} \textrm{with} \hspace{1cm }\delta = 3,
\end{equation}
where

\begin{equation}
    S^{g0}  = \frac{gD^2F(\theta)\Lambda k_B^4 4! \zeta (4)}{2^{7/2} \pi^{3/2} e \rho_m \hbar^4\nu_s^4 {\nu_F}^{*2}n_s^{1/2}}. 
\end{equation}
Hence, in the BG regime $S_g$ $\sim$ ${\nu_F}^{* -2}$ ,  $T^3$,  and $n_s^{1/2}$.  Moreover, $S_g$ $\sim$ $\nu_s^{-4}$, indicating that even a small change in $\nu_s$ will result in to a large change in $S^g$. The BG regime relation can be used to determine  the el-ap coupling constant $D$ by experimentally studying $S^g$ as a function of temperature.

\subsection{Diffusion thermopower }
The formula for  diffusion thermopower $S^d$  in  tBLG can be obtained from Boltzmann transport equation in the relaxation time approximation, following \cite{38}. It is given by
\begin{equation}
S^d = - \frac{1}{eT} [ \{\frac{\langle \tau (E_k) E_k \rangle} {\langle\tau(E_k)\rangle} -E_F^*\}], 	
\end{equation}                 
where
\begin{equation*}
    \langle x \rangle = \frac{[ \int x (-\partial f/\partial E_k)D(E_k)dE_k] }{[\int (-\partial f/\partial E_k)D(E_k)dE_k]}. 
\end{equation*}

In the low  $T$ region and for large $n_s$, $S^d$ is given by the Mott formula \cite{23}, taking the  energy dependence of $\tau(E_k) = \tau_0 E_k^s$
\begin{equation}
     S^d  = -[\pi^2k_B^2T(s +1)/3eE_F^*], 	 \end{equation}
where $s$  $\sim |1|$  is the exponent of energy. 

Since we need to use the acoustic phonon limited resistivity in tBLG, in the calculations of power factor, we give here  its equation  from the Refs. \cite{10,13}
\begin{equation}
    \rho = \frac{32D^2F(\theta)k_F^*}{ge^2\rho_m \nu_F^{*2}\nu_s}I(T/T_{BG}),
\end{equation}
where
\begin{equation}
    I(z) = \frac{1}{z} \int_0^1 dx x^4(1-x^2)^{1/2} \frac{e^{x/z}}{(e^{x/z} - 1)^2}.
\end{equation}				                               
This is the equation which is obtained by appropriate modification of the equation  (2) of \cite{39} for MLG. 

\section{Results and discussion}
In this section we have numerically  evaluated and discussed the results for phonon-drag thermopower $S^g$, diffusion thermopower $S^d$ and power factor $PF$ as a function of twist angle $\theta$ ( 1.1$^{\circ}$, 1.2$^\circ$ and 1.3$^\circ$),  temperature $T$ (1-20 K), and  electron density $n_s$(0.5- 1 $n_0$) with $n_0 =1\times10^{12}$ cm$^{-2}$. The material parameters known for graphene are used \cite{10,13}: $\rho_m = 7.6\times10^{-8}$ g/cm$^2$, $\nu_F = 1\times10^8$ cm/s, $\nu_s = 2\times10^6$ cm/s, $D$ = 20 eV, $\theta_m =1.02^{\circ}$ and $\Lambda = 10\mu$m \cite{10,36}. Our choice of $D$ here is the one used by many research groups to explain their transport data in MLG \cite{33,34,35,40,41}. In order to see the large influence of the twist angles the chosen  1.1$^\circ$, 1.2$^\circ$ and 1.3$^\circ$ are closer to the  magic angle. The corresponding effective Fermi velocities ${\nu_F}^*$ = 4, 9 and 14$\times 10^6$ cm/s ($>$ 1.5 $v_s$ \cite{13}) are much smaller than the bare Fermi velocity $v_F$. The values of the function $F(\theta$) for these angles are taken from figure 3 of Das Sarma et al \cite{13}. The range of $T$ and $n_s$ chosen are such that the system remains degenerate (as required in the present theory), slightly away from the Dirac point, and within the linear region of moire flat band. For $T < \sim$ 1 K, the tBLG is likely to be in superconducting state. For the temperature region considered, in the $S^g$ calculation, the phonon relaxation time is assumed to be due to the boundary scattering \cite{36}. We have set the specular parameter, which is a  measure of the roughness at the graphene edge, $p = 0$ with a reasonable choice of  $\Lambda$ \cite{23,36}. It is to be noted that $\Lambda$ and $ p (0 \leq  p \leq 1$) are sample dependent.

\begin{figure}[h]
\centering
\includegraphics[angle=-0.0,origin=c,height=7.6cm,width=8.7cm]{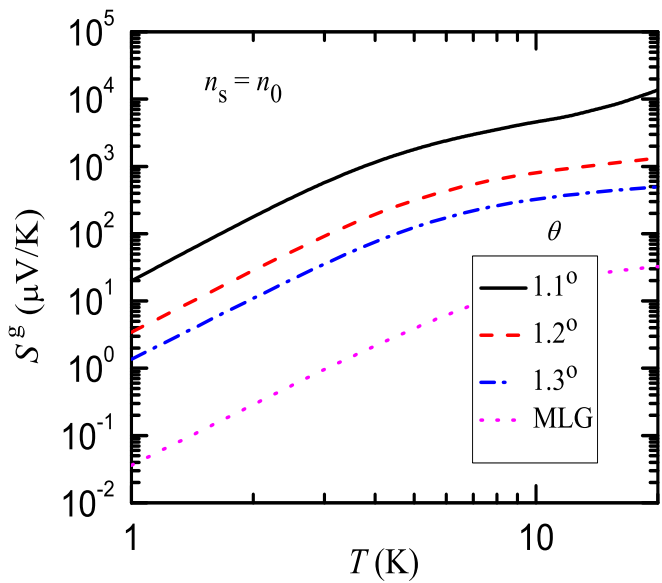}
\caption{Phonon-drag thermopower $S^g$ as a function of temperature
$T$ for different twist angles $\theta$ in tBLG and for MLG with $n_s = n_0$ .}. 
\label{fig1}
\end{figure}

To investigate the effect of twist angle on phonon-drag thermopower in tBLG, we have shown in figure \ref{fig1} $S^g$ as a function of $T$ for $\theta$ = 1.1$^\circ$, 1.2$^\circ$ and 1.3$^\circ$ for electron density $n_s$=  $n_0$ along with the corresponding curve for MLG obtained from Ref. \cite{23}. The behavior of $S^g$ is generic with rapid increase in the low temperature region and slowly increasing in the higher temperature region and finally flattening with a knee like structure. At very low temperature all the curves are expected to show $S^g$ $\sim$ $T^3$ power law behavior, as expected in the BG regime ($T << T_{BG}$), and it is  manifestation of 2D phonons. In our calculations it is obeyed for for $T  <$  1 K for  $\theta$ = 1.1$^\circ$, 1.2$^\circ$  and up to $\sim$1 K for $\theta$ = 1.3$^\circ$ and hence  not observed in figure \ref{fig1} for the curves corresponding to tBLG. It is apparent that, in tBLG, the range of $T$ for this power law is  $\theta$  dependent, and it is smaller for smaller $\theta$, as found in hot electron power loss \cite{15}. On the other hand, in MLG this power law is satisfied up to $\sim$ 2.5 K. Expressing $S^g$ $\sim$ $T^{\delta}$, the exponent $\delta$, which depends upon the electron- phonon interaction,  decreases from 3 to nearly zero as $T$ increases. In the case of the curve for $\theta$ =1.1$^\circ$, we notice a kind of upward trend of $S^g$ for about $T > \sim$ 15 K. This may be attributed to the less validity of our model, which is for the degenerate system and with the increasing $T$ the temperature dependence of the chemical potential $\mu$ moves too close to the Dirac point for this $\theta$.

Our very important observation is that $S^g$ in tBLG is strongly twist angle dependent. It is found that,  compared to MLG, $S^g$ is highly enhanced in tBLG,    attributing to the strongly suppressed effective Fermi velocity ${\nu_F}^{*}$ because of the twist angle. Since, in BG regime, $S^g \sim {\nu_F}^{*-2}$, $S^g$ (tBLG ) / $S^g(MLG) \sim (\nu_F/ {\nu_F}^*)^2$. For $\theta$ = 1.1$^\circ$, 1.2$^\circ$ and 1.3$^\circ$, the ratio ($\nu_F/ {\nu_F}^*)^2 = 625$, 123.5, and 51.0, respectively. Consequently, in BG regime, the dramatic effect is that, the ratio ($v_F/ {\nu_F}^*)^2$ expected to produce the enhancement of $S^g$ in tBLG  by these factors, respectively, for $\theta$ = 1.1$^\circ$, 1.2$^\circ$ and 1.3$^\circ$. The enhancement is highest for  the $\theta$  closest to $\theta_m$ and  as $\theta$ increases  $S^g$  is decreasing and finally approaching the value closer to that of MLG for $\theta = 3.0^{\circ}$. Moreover, there are other factors such as $g_l$, $F(\theta$) and $k_F^*$  which also make the difference (by a factor of $\sim$ 0.7) in the magnitudes of   $S^g$ (tBLG ) with $S^g$ (MLG). With increasing temperature, it is observed from our calculations that,  for a given $\theta$, the magnitude of enhancement of $S^g$ in tBLG decreases. For example, for $\theta$ = 1.1$^\circ$,  at $T$ = 1, 5 and 10 K, the  ratio [$S^g$ (tBLG) / $S^g$ (MLG)] = 552.63,   456.96 and 272.78, respectively. These  ratios, respectively,  for 1.2$^\circ$  (1.3$^\circ$) are 94.74 (37.67),  80.38 (31.65), 47.633 (19.35). We would like to emphasize that  in tBLG the twist angle acts as an additional  strong tunable parameter of $S^g$, similar to the observation made with respect to linear-in-$T$ resistivity in tBLG \cite{10,13} and hot electron power loss \cite{15}. In the present calculations, the predicted largest $S^g$ is about 15 mV/K at 20 K for $\theta$ = 1.1$^\circ$.

In the BG regime, we find the  following important and useful relations in tBLG. Since the underlying mechanism for phonon limited mobility $\mu_p$, phonon-drag thermopower $S^g$ and the hot electron power loss $F_e(T)$ is the same, relations between them are expected. They are established in conventional 2DEG and 3DEG \cite{19,20,42,43,44} and 2D and 3D Dirac fermions \cite{23,45,46}. The relation between $\mu_p$ and $S^g$ is a well known Herring’s law $S^g\mu_p= - (\Lambda \nu_s/T)$ \cite {42}. In the following we shall find these relations in tBLG. In BG regime, by using the mobility equation $\mu_p(T) = [(15 ge \hbar^4\rho_m \nu_s^5{\nu_F}^{*2} n_s^{1/2})T^{-4}]/[16 \sqrt{2} \pi^{5/2}D^2k_B^4 F(\theta)]$ \cite{15}, derived from the resistivity  equation of Ref. \cite{10}, with our Eq. (3) for $S^g$,  we find  $S^g \mu_p= - (\Lambda v_s/T)$ and hence Herring’s law is satisfied in tBLG. This relation is the same as found in conventional 2DEG  \cite{19, 20, 43} and in MLG \cite{23}. $S^g$ is also related to $F_e(T)$, obtained in Ref \cite{15}, by a simple equation $S^g  = (2\Lambda / ev_sT) F_e(T)$, which is again the same as found for 2DEG in MOSFET  \cite{43} and Dirac fermions in MLG \cite{23}. These relations are significant due to the fact that, if one of them is measured, the other one can be calculated. For example, phonon-limited mobility of electrons and composite fermions was determined from the $S^g$ measurements in GaAs heterojucntions  \cite{47}.

\begin{figure}[h]
\centering
\includegraphics[angle=-0.0,origin=c,height=7.6cm,width=8.7cm]{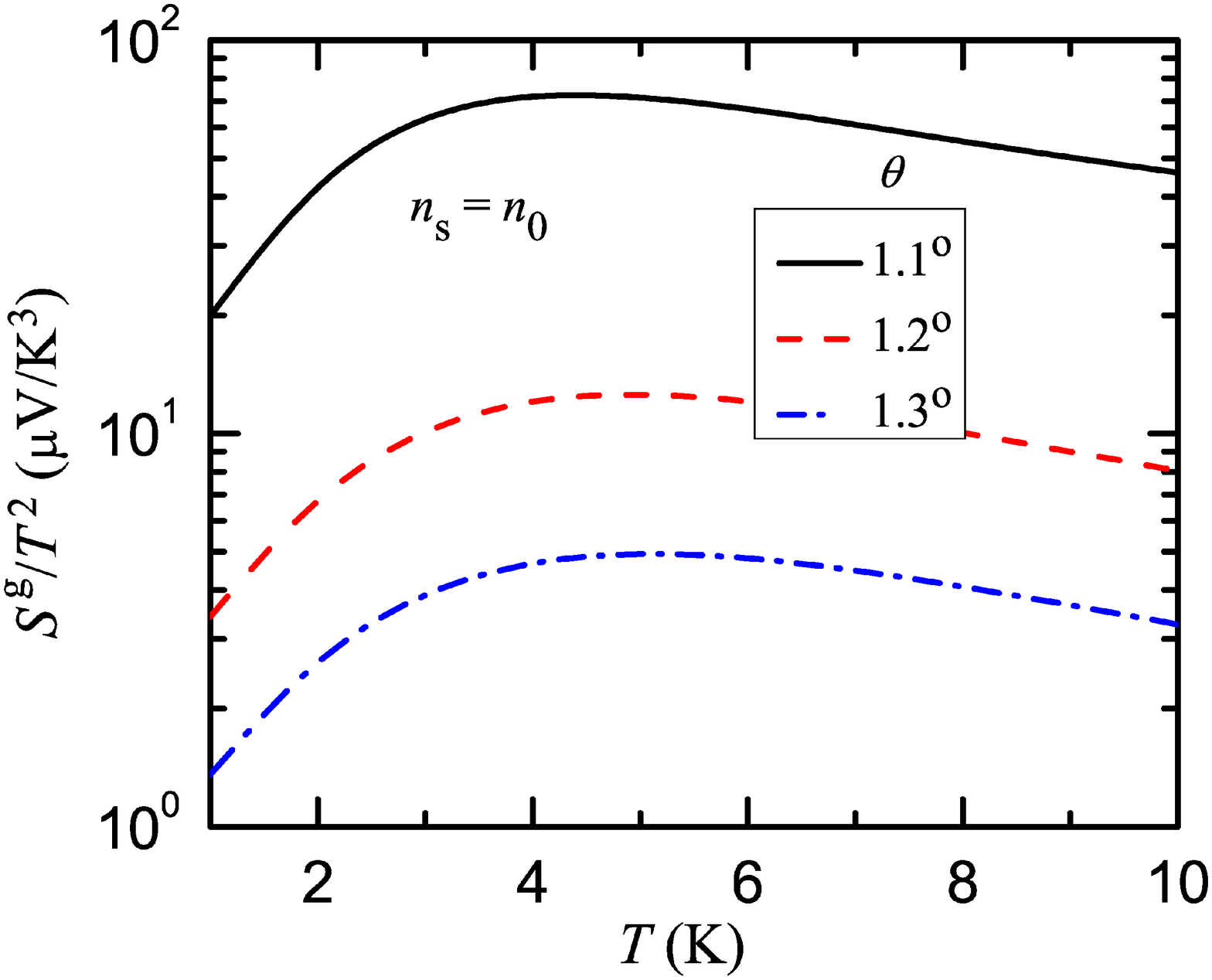}
\caption{$S^g/T^2$ as a function of temperature $T$ for different twist
angles $\theta$ in tBLG with $n_s = n_0$}. 
\label{fig2}
\end{figure}

In the following we will find the temperature at which the maximum momentum transfer takes place, in the phonon-drag,  from phonons to the electron system. From the simple balance approach, $S^g \alpha -f (n_s, T) C_v(T)/(n_s e)$ \cite{19}, where  $f(n_s, T)$  is the fraction of the momentum transferred from phonons to electrons and  $C_v(T)$ is the lattice specific heat which is $\sim T^2$  at low temperature in 2D systems. The low - $T$ lattice thermal conductivity also varies as $\sim T^2$ in the boundary scattering regime. By plotting $S^g/T^2$ as a function of $T$, the temperature $T_{KA}$ corresponding to a maximum of $S^g/T^2$ gives us the temperature at which phonons transfer maximum momentum to the electrons. This ‘Kohn anomaly temperature $T_{KA}$' is defined by $2\hbar \nu_sk_F^* = Ck_BT_{KA}$ for which the dominant phonon wave vector  $q_D = Ck_BT/\hbar \nu_s$ \cite{19, 20} for $q_D = 2k_F^*$. Here $C$ is a dimensionless constant which we  found to be 9.0 in tBLG. In figure \ref{fig2}, we have depicted $S^g/T^2$  as a function of $T$ for $\theta$ = 1.1$^\circ$ ,1.2$^\circ$ and 1.3$^\circ$ with $n_s= n_0$. For each  $\theta$, curve shows a maximum at $T_{KA}$ which is greater for larger $\theta$ and we find a relation that ($T_{KA}$ / $\theta$) = constant $\sim$ 4. The $T_{KA}$ dependence on $\theta$ is surprising and a new result in tBLG, although by definition $T_{KA}$ depends upon only $k_F^*$ and hence $n_s$. This also supplements the observation made in figure \ref{fig1} regarding the range of validity of $T^3$ power law for different $\theta$. The rapid rise at low $T$ may be attributed to the increasing number of phonons as their dominant wave vector  $q_D \sim k_BT$ \cite{19,20}. For $T > T_{KA}$, electrons couple to fewer phonons, as phonons with $q = 2k_F^*$ are less excited, causing the decrease of $S^g /T^2$.

\begin{figure}[h]
\centering
\includegraphics[angle=-0.0,origin=c,height=7.6cm,width=8.7cm]{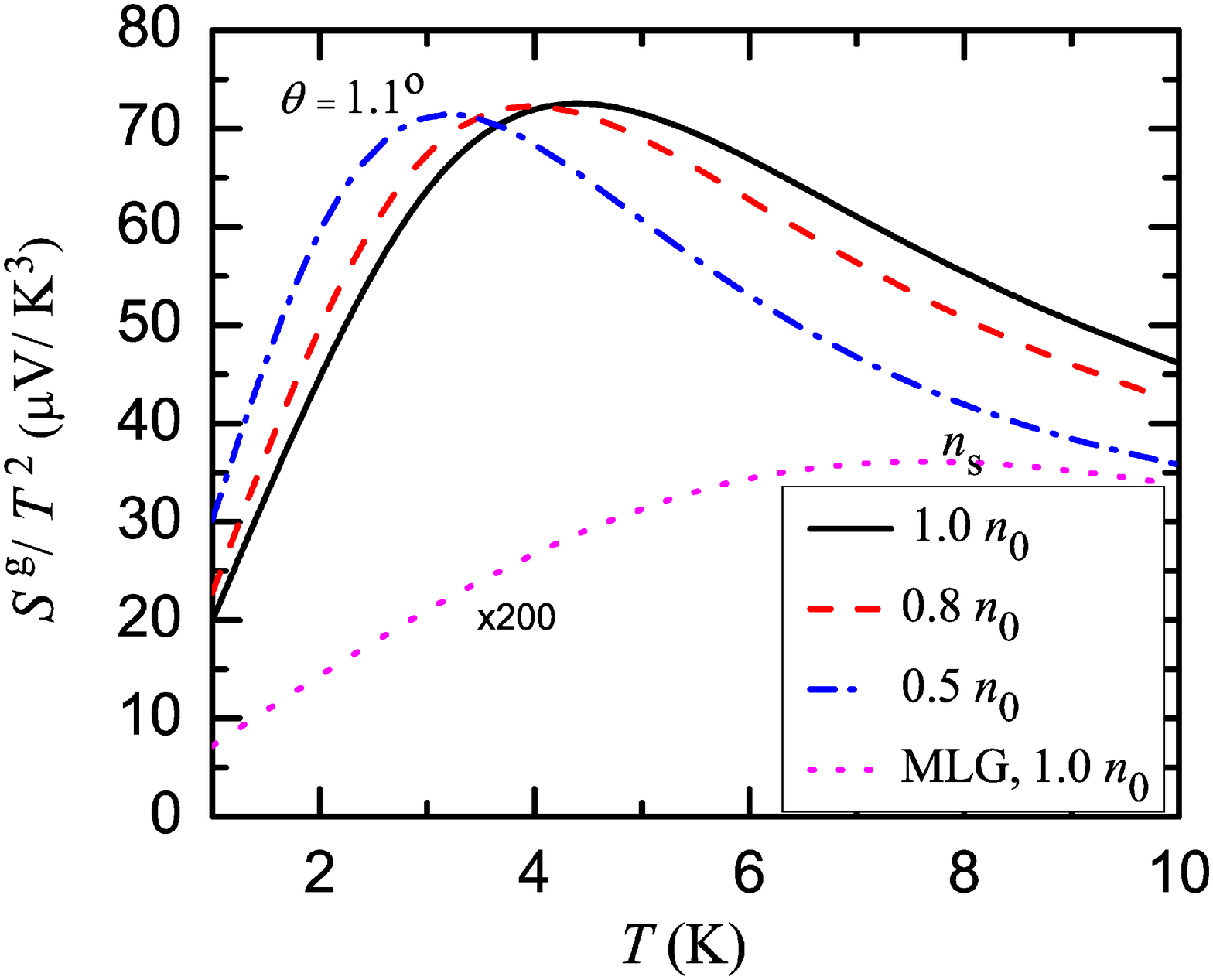}
\caption{$S^g/T^2$ as a function of temperature $T$ for different electron
densities $n_s$ in tBLG for $\theta$ = 1.1$^\circ$. The same is plotted for MLG, with
magnification by 200, for $n_s = n_0$ .}. 
\label{fig3}
\end{figure}

In order to investigate the electron density dependence, $S^g / T^2$ vs $T$ is plotted in figure \ref{fig3} for $n_s$= 0.5, 0.8 and 1.0 $n_0$ taking $\theta$ =1.1$^\circ$. The curve corresponding to MLG is for $n_s= n_0$ and it is shown with the magnification by a factor of 2$\times 10^2$. The $T$ dependence of each of the  curves is same as in figure \ref{fig2}, with a maximum occurring at $T = T_{KA}$. In this case, $T_{KA}$ is shifting to higher value for larger $n_s$, resulting in to a relation ($T_{KA}$ /$\sqrt{n_s}$) = 4.25$\times 10^{-6}$ K-cm, which is understandable by definition of $T_{KA}$. Consequently, $T_{KA} =$ 0.11  $T_{BG}$. It also indicates that $T^3$ power law is satisfied for larger range of $T$ for greater $n_s$. At low temperature $S^g$ is greater for smaller $n_s$, say upto $T \sim T_{KA}$, where as it is higher for larger $n_s$ for $T>\sim T_{KA}$. In BG regime, $S^g \sim  n_s^{-1/2}$ dependence is attributed to the 2D nature of Dirac fermions, as found in MLG \cite{23},   in contrast to the $n_s^{-3/2}$ behavior in conventional 2DEG  \cite{43}.

At this point we would like to address, for comparison, the dominant phonon wave  vector $q_D = Ck_BT/\hbar v_s$ in the electron-phonon scattering rate, in different systems. For conventional 2DEG in Si-MOSFET and GaAs heterojucntions, the value of $C$ is shown to be  $\sim$ 5 \cite{19,20}. By revisiting the  $S^g$ / $T^2$ vs $T$ curves in MLG \cite{23}, monolayer of molybdenum disulphide (MoS$_2$)  \cite{48} and three-dimensional Dirac semimetals (3DDS)  \cite{45}, from the  position of maximum in each of these systems, we determine $C$ = 7.0,  5.6  and 8.6, respectively.
 
 The diffusion thermopower, which is sensitive measure of the energy dependence of the electron momentum relaxation time, for the degenerate electron gas at low temperature is given by the Mott formula (Eq. (6)). It is shown that for acoustic phonon scattering, which is the only known  resistive source in tBLG samples \cite{9,10,11,13}, $s = -1$ for which $S^d = 0$. The nature of the impurity scattering in tBLG is not precisely known, although there exist early calculations of impurity induced resistivity \cite{14}. In view of this, at this initial stage, we have presented a representative calculation of $S^d$ from the Mott formula, for simplicity, taking $s = 1$ corresponding to the unscreened impurity scattering in MLG \cite{24}. Consideration of the detailed energy dependence of $\tau(E_k)$, in MLG, has reduced $S^d$ by a factor of $\sim$ 2 with parabolic temperature dependence \cite{24}. We also point out that, in the temperature region where acoustic phonon scattering is the dominant resistive mechanism, applying the Matthiessen’s rule, there may not be contribution of $S^d$ from the resultant relaxation time. It may be noted that in Si-inversion layer, $n_s$ was tuned to make $s = -1$ so that $S^d = 0$ and the measured thermopower was only due to $S^g$  \cite{43}. 

\begin{figure}[h]
\centering
\includegraphics[angle=-0.0,origin=c,height=7.6cm,width=8.7cm]{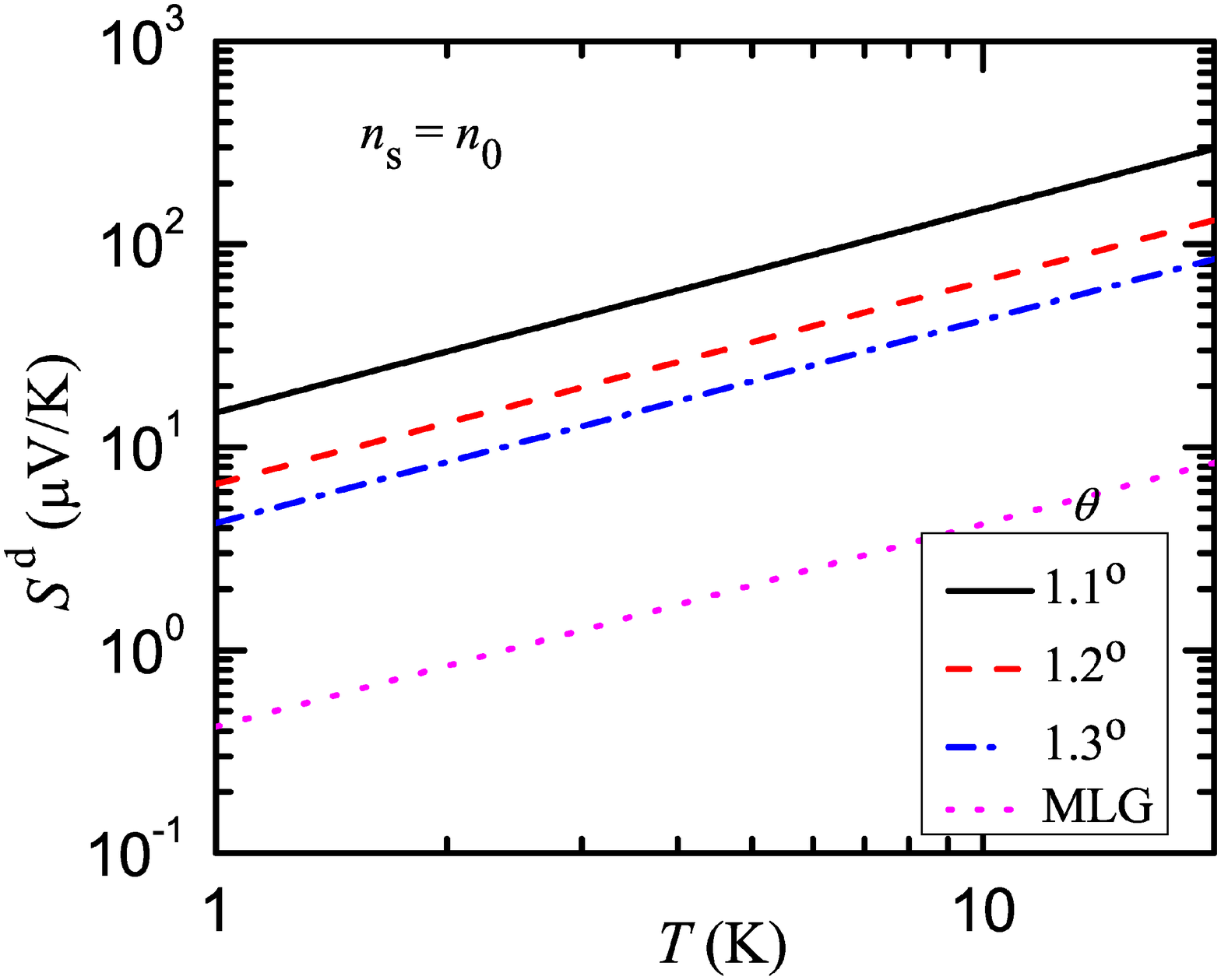}
\caption{Diffusion thermopower $S^d$ as a function of temperature $T$
for different twist angles $\theta$ in tBLG and for MLG with $n_s = n_0$}. 
\label{fig4}
\end{figure}

In order to see the effect of twist angle, $S^d$ is plotted as a function of $T$ for different $\theta$ in tBLG along with the $S^d$ in MLG in figure \ref{fig4} taking $n_s = n_0$. It is found that there is a big enhancement of $S^d$ compared to that in MLG. For example, at $T = 1$, 5 and 10 K, respectively, the enhancement is by 35, 15 and 10 times. The enhancement is smaller for larger $\theta$. This is, as expected, due to suppressed Fermi energy, which is $\theta$ dependent, in tBLG. Hence, $\theta$ is another important tunable parameter of $S^d$ in tBLG. However, enhancement of the $S^d$ is much smaller than that of $S^g$.

\begin{figure}[h]
\centering
\includegraphics[angle=-0.0,origin=c,height=7.6cm,width=8.7cm]{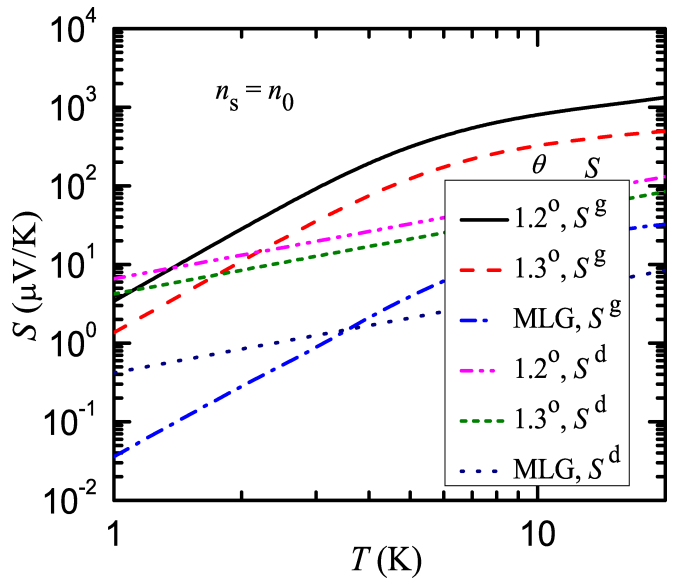}
\caption{$S^g$ and $S^d$ as a function of temperature $T$ for different
twist angles $\theta$ in tBLG and for MLG with $n_s = n_0$.} 
\label{fig5}
\end{figure}

With an objective of finding the temperature region in which either $S^g$ or $S^d$ is dominant, we have shown both of them as a function of $T$ for $\theta$ =1.2$^\circ$ and 1.3$^\circ$ with the respective curves of MLG in figure \ref{fig5}. We observe that there is a crossover of $S^g$ and $S^d$ at a temperature $T_c$, above which $S^g$ is dominating, and this $T_c$ is $\theta$ dependent. For $\theta$ =1.2$^\circ$ (1.3$^\circ$) the cross over takes place at $\sim$ 1.3 (1.8) K in tBLG and $T_c$ = $\sim$ 3.5 K in MLG. $S^d$ is found to be dominant in a small region of low $T$.

We would like to make a qualitative comparison of $S$ of our calculations with the out-of-plane $S$ measured in tBLG \cite{33_1} and theoretically estimated in tBGNR \cite{35_1}. In tBLG, the measurements of $S$ are over the range of temperature $T$ = 20-300 K and for specific $\theta$’s \cite{33_1}. For large $\theta$ ($\sim$ 12.5$^{\circ}$), the authors analyze the measured $S$ by the out-of-plane phonon-drag $S^g$, due to the LBM phonon modes of quadratic dispersion, over the entire temperature range, with a maximum of $\sim$ 35 $\mu$V/K for $T > \sim$ 100 K, which shifts to higher $T$ for larger $n_s$. This measured $S$ is found to be smaller for larger $n_s$ (i.e. $S = S^g$ $\sim {n_s}^{-1}$ in Eq.(1) of Ref.\cite{32_1}). In contrast, for small $\theta$ ($< 6^{\circ}$), measured $S$ is shown to exhibit Mott thermopower ($S^d \sim k_BT/E_F$) in the experimental temperature range $\sim$ 30-300 K, with a value of $\sim$ 30 $\mu$V/K at 300 K. However, in their out-of-plane theoretical analysis of either phonon-drag or diffusion thermopower, we don’t notice any twist angle dependence. On the other hand, our in-plane calculated $S^g$, for $\theta$ (=1.1$^\circ$ - 1.3$^\circ$) close to $\theta_m$, at about $\sim$ 10 K, is about $\sim$ 10$^3$ - 10$^4$ times larger than these measured ones extrapolated to 10 K. Moreover, for larger $n_s$, $S^g$ is smaller (larger) at $T < T_{KA}$ ($T > T_{KA}$). Besides, our calculated $S^g >> S^d$ for $T > \sim$ 2 K, and both $S^g$ and $S^d$ are very much $\theta$ dependent through ${\nu_F}^*$. We point out that our in-plane theory, for $\theta$’s close to $\theta_m$, cannot be applied to interpret the across plane experimental data of Ref. \cite{33_1}. In tBGNR, by the first principles calculation, with Landauer-Buttiker and Boltzmann theories, a peak value of $S \sim$ 0.75 mV/K is predicted for twist angles 21.8$^\circ$ and 38.2$^\circ$ at 300 K \cite{35_1}. This value is closer to our predicted $S = \sim$ 1-10 mV/K at $T \sim$ 10 K and for $\theta$’s closer to magic angle.

\begin{figure}[h]
\centering
\includegraphics[angle=-0.0,origin=c,height=7.6cm,width=8.7cm]{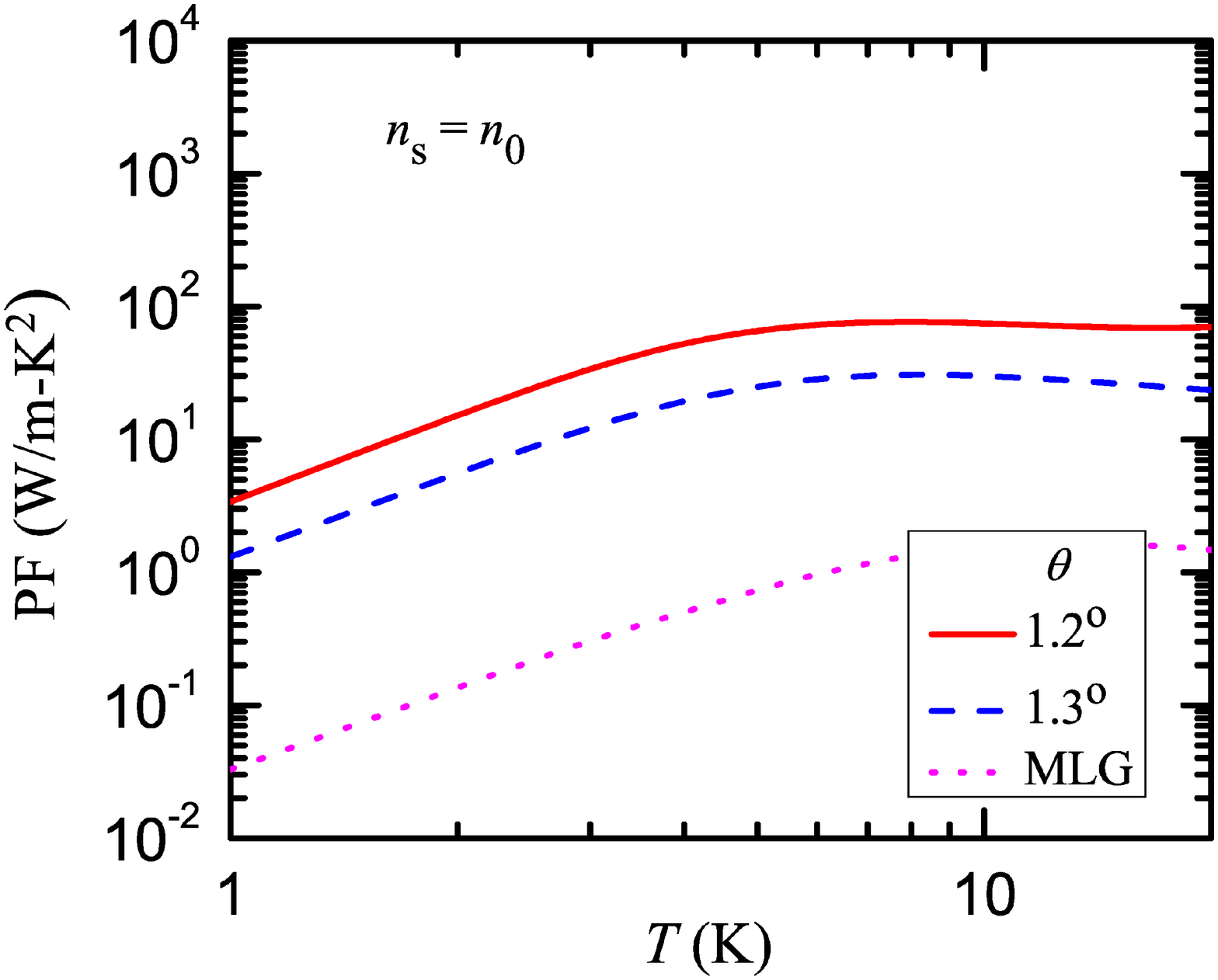}
\caption{Power factor $PF$ as a function of temperature $T$ for different
twist angles $\theta$ in tBLG and for MLG with $n_s = n_0$.} 
\label{fig6}
\end{figure}

In figure \ref{fig6}, the power factor is shown as a function of temperature for $\theta$ =1.2$^\circ$ and 1.3$^\circ$. In the $PF = (S^2/\rho t)$ calculation, with  the thickness $t$ = 0.35 nm \cite{37}, we have taken only $S^g$ contribution, because  $S^d << S^g$ in the larger range of $T$ considered, besides it is likely to be nearly zero in the region where acoustic phonon scattering is limiting the transport. The required resistivity is calculated using Eq (7) taken from the Ref. \cite{10}. The $PF$ calculations of MLG are obtained using equation of $S^g$ from the Ref. \cite{23} and $\rho$ of Ref.  \cite{39}. $PF$ is found to increase with increase of $T$ in low $T$ region, then slowly  saturates and marginally declines in the high $T$ region. The temperature of near saturation / decline shifts to the  higher temperature as $\theta$ increases. Compared to MLG, we observe a giant enhancement of $PF$ and this enhancement is again $\theta$ and $T$ dependent, being largest for smallest $\theta$ and  $T$. For example for $\theta$ = 1.2$^\circ$, the enhancement of $PF$ is by a factor $\sim$ 103 at 1 K and for $\theta$ = 1.3$^\circ$ it is about 16 times at 20 K. Again, twist angle is found to be a very significant tuning parameter of the power factor, besides temperature and electron density. In the present work, for $\theta$ = 1.2$^\circ$ the highest power factor is found to be $\sim$ 75 W/m-K$^2$ at about 8 K. We point out that, it is an unique situation in tBLG in which  the large enhancement of $S \sim S^g$ is nearly equal to the suppression of $\sigma$, due to the strongly increased el-ap coupling, and hence the giant $PF$.

We roughly estimate the figure of merit $ZT$ in the case of tBLG at $\sim$ 20 K for $\theta$ = 1.2$^{\circ}$. Since, in tBLG it is assumed that acoustic phonon  spectrum is not affected by the twist angle \cite{10}, we can take the phonon contribution $\kappa_p$ to the total thermal conductivity $\kappa$ =  $\kappa_p$ +  $\kappa_e$  to  be the same as in MLG. Moreover, in tBLG electronic contribution $\kappa_e$  is expected to be very much suppressed  due to the  strongly reduced  ${\nu_F}^*$  because  by Wiedemann–Franz law $\kappa_e$  $\alpha$ $\sigma$ and hence we take $\kappa \simeq \kappa_p$ .  In order to estimate $ZT$, we chose the measured value  $\kappa$ = $\sim$10 W/m-K at $\sim$ 20 K  \cite{49}. Then, with $PF= \sim$70 W/m-K$^2$, using $ZT = (PF\times T)/ \kappa$,  we get $ZT$ =140, an extremely large value.

We would like to make the following remarks. It is to be noted that there is some uncertainty in the value of el-ph coupling constant $D$ (=10- 40 eV) in the literature  \cite{10,23,50} and any change in its value  will change $S^g$ significantly as $S^g$ $\alpha$ $D^2$. Moreover, the strong dependence of $S^g$  on the phonon velocity $v_s$ (approximately $S^g \sim v_s^{-4}$) demands an accurate knowledge of $v_s$.  Secondly, an ideal expression for the phonon relaxation time is given by  $\tau_p = (\Lambda/v_s)[(1+p)/(1-p)]$, where $p$ is the specularity parameter. Nika et al  \cite{36}  have used $p = 0.9$ in order to explain thermal conductivity data in MLG, and this choice  will increase the effective  mean free path  and hence $S^g$ by nearly 20 times. 
 \section{Summary}
The in-plane thermopower and power factor are calculated theoretically in tBLG showing the profound effect of suppression of the effective Fermi velocity ${\nu_F}^*$, due to the twist angle $\theta$, in moiré flat band. Our minimal naïve theory is analytical and is of more relevant in the early stage of the  development of the subject, to bring out essential features of these transport  properties. Our simple model is for $n_s \leq 10^{12}$ cm$^{-2}$ and with the chemical potential  little away from the Dirac point. 

The $\theta$ dependent ${\nu_F}^*$  is found to strongly enhance the phonon-drag thermopower $S^g$ and diffusion thermopower $S^d$, for small $\theta$ closer to $\theta_m$. This effect is more on $S^g$ due to the highly increased electron- acoustic phonon interaction, which is also the source for the observed  large linear-in-$T$ resistivity \cite{10}. The enhancement of $S^g$ up to $\sim$ 500 times that in MLG is predicted with a large and measurable maximum value of $\sim$ 10 mV/K at about 20 K; and is found to significantly dominate over $S^d$ for large range of $T$($ >\sim$2 K). In tBLG, the twist angle  has emerged as a strong tunable parameter of thermopower, besides $T$ and $n_s$. The effect of ${\nu_F}^*$ on $S^g$ and $S^d$ is found to reduce with the increasing $\theta$ and $T$. In the BG regime, power laws $S^g \sim {\nu_F}^{* -2}$, $T^3$ and $n_s^{-1/2}$ are obtained. The temperature range of validity of $T^3$ law is also found to be $\theta$ dependent. In the temperature range considered, the $T$ dependence of $S^g$ is generic with rapid increase at low $T$ and nearly constant at higher $T$. On the contrary, $S^d$ is taken to be governed by the simple  Mott formula with $S^d$ $\sim$ ${\nu_F}^{*-1}$, $T$  and $n_s^{-1/2}$. By plotting the $S^g/T^2$  vs $T$, for different $\theta$ and $n_s$, we have found simple relations of 'Kohn anomaly temperature $T_{KA}$'  with $\theta$,  $n_s$ and $T_{BG}$, namely $T_{KA}$/ $\theta$ = constant ($\sim$ 4) and $T_{KA}$/$ \sqrt{n_s} = $constant ($\sim$ 4.25 $\times$ 10 $^{-6}$ K-cm) and $T_{KA}$/$T_{BG}$= 0.11. In the BG regime, simple relations of $S^g$ with phonon limited mobility $\mu_p$ i.e $S^g \mu_p= - \Lambda \nu_s/T$ (Herring’s law) and the hot electron power loss $F_e(T)$  i.e  $S^g  = (2\Lambda / e\nu_sT) F_e(T)$  are obtained, and these are useful in determining the one by measuring the other. Our calculations of $S$ are qualitatively compared with the measured values across the plane of tBLG \cite{33_1}. 
Power factor $PF$ is also investigated as a function of $\theta$ and $T$. The $PF$ is found to be highly enhanced for $\theta$ closer to $\theta_m$, thereby   $\theta$ acting as its  strong tunable parameter. $PF$ as a function of $T$ exhibits a broad maximum and a maximum value of $PF \sim$ 75 W/m-K$^2$ is predicted for $\theta$ =1.2$^\circ$. As a result, it is discussed that an extremely large figure of merit is possible. We believe, our findings with this simple analytical model will
initiate some experimental work on the in-plane thermopower in tBLG and be useful to explain the experimental data.\\\\
\textbf{Acknowledgement:} Author wishes to acknowledge Vidyashree Hebbar and Ravi Kashikar for their help in preparing this manuscript in Latex.

\bibliography{paper}%

\providecommand{\noopsort}[1]{}\providecommand{\singleletter}[1]{#1}%

\end{document}